\documentclass[11pt]{article}
\usepackage{hyperref}
\pdfoutput=1

\begin{document}
\title{Can mosquitoes fly in the rain?}
\author{Andrew Dickerson, Peter Shankles, Nihar Madhavan, David Hu \\
\\\vspace{6pt} School of Mechanical Engineering, \\ Georgia Institute of Technology, Atlanta, GA 30332, USA}
\maketitle
\begin{abstract}
Collisions with raindrops are one of many obstacles insects face during flight. In this fluid dynamics video, we present a series of high-speed films of impacts between mosquitoes and raindrops.   We also present drop impacts upon insect mimics, which are unsupported styrofoam balls of the same mass as mosquitoes. High-speed videography and particle tracking during collision are employed to determine the insect position versus time. We determine the magnitude of acceleration by considering the momentum transfer and impact duration. Experiments with live mosquitoes indicate a surprising ability to quickly recover flight post-collision, despite accelerations of 30-300 gravities over durations of 1 ms.

\end{abstract}
%
\end{document}